\documentstyle[a4,11pt] {article}
\begin{document}

\setlength{\baselineskip}{20pt}

\noindent
{\Large\bf Comment on ``Dispersion relation for MHD waves in homogeneous 
plasma''}

\vspace{4mm}

\setlength{\baselineskip}{15pt}

\noindent
{\bf Suresh Chandra, G.M. Dak, B.K. Kumthekar and Monika Sharma}

\vspace{2mm}

\noindent
School of Physics, Shri Mata Vaishno Devi University, Katra 182 320 (J\&K), 
India

\vspace{2mm}

\noindent
Email: suresh492000@yahoo.co.in

\vspace{6mm}

\noindent
{\bf Abstract:}
Pandey \& Dwivedi (2007) again tried to claim that the dispersion relation
for the given set of equations must be a sixth degree polynomial. Through a 
series of papers, they are unnecessarily creating confusion. In the 
present communication, we have shown how Pandey \& Dwivedi (2007) are 
introducing an additional root, which is insignificant. Moreover, five roots of
 both the polynomials are common and they are sufficient for the discussion of
 propagation of slow-mode and fast-mode waves. 

\vspace{2mm}


\section{Introduction}

For application of magnetohydrodynamics (MHD) in solar physics as well as in
plasma physics, dispersion relation plays key role. The basic equations under 
the present investigation can be expressed as (Pandey \& 
Dwivedi, 2007, hereinafter referred to as PD)
\begin{eqnarray}
\frac{\partial \rho}{\partial t} +\nabla . \ (\rho \hspace{-0.5mm}
\stackrel{\rightarrow}{v} ) = 0 \hspace{4.9cm}\label{01}  \\ 
\rho \frac{D \hspace{-1mm} \stackrel{\rightarrow}{v}}{D t} =
 -\nabla p + \frac{1}{4 \pi}(\nabla \times \stackrel{\rightarrow}{B})\times
\stackrel{\rightarrow}{B} - \nabla . \Pi \hspace{1.9cm} \label{02} \\ 
\frac{\partial  \hspace{-1mm} \stackrel{\rightarrow}{B}}{\partial t} = \nabla
\times (\stackrel{\rightarrow}{v} \times \stackrel{\rightarrow}{B})
 \hspace{4.7cm} \label{03} \\ 
\frac{D p}{D t} + \gamma p (\nabla . \stackrel{\rightarrow}{v}) = (\gamma -1)
\Big[\nabla . \ \kappa \nabla T + Q_{vis} - Q_{rad}\Big] \hspace{-0.4cm} \label{04} \\ 
 p = \frac{2\rho k_B T}{m_p} \hspace{6.1cm}  \label{05}
\end{eqnarray}

\noindent
Here, symbols have their usual meaning. The quantities $Q_{th}$, $Q_{vis}$ and 
$Q_{rad}$ are 
\begin{eqnarray}
Q_{th} = \kappa_\parallel \Big(\frac{\partial T}{\partial z}\Big)^2 T^{-1} 
\hspace{2cm} Q_{vis} = \frac{\eta_0}{3} (\nabla . \stackrel{\rightarrow}{v})^2
\hspace{2cm} Q_{rad} = n_e n_H Q(T) \nonumber
\end{eqnarray}

\noindent
where $\kappa_\parallel$ represents the conductivity along the magnetic field
and is expressed by $\kappa_\parallel \approx  10^{-6} T^{5/2}$. For small 
perturbations from the equilibrium, we have
\begin{eqnarray}
\rho = \rho_0 + \rho_1 \hspace{2.5cm} \stackrel{\rightarrow}{v} = \stackrel{
\rightarrow}{v}_1 \hspace{3.5cm} \stackrel{\rightarrow}{B} = \stackrel{
\rightarrow }{B}_0 + \stackrel{\rightarrow}{B}_1 \nonumber\\ 
p = p_0 + p_1 \hspace{2.7cm} T = T_0 + T_1 \hspace{2.6cm} \Pi = \Pi_0 + \Pi_1
  \nonumber
\end{eqnarray}

\noindent
where the equilibrium part is denoted by the subscript `0' and the perturbation 
part by the subscript `1'. For the magnetic field taken along the $z$-axis, 
({\it i.e.,} $\stackrel{\rightarrow}{B}_0 = B_0 \hat{z}$) and the propagation 
vector $\stackrel{\rightarrow}{k} = k_x \hat{x} + k_z \hat{z}$, the equations 
(\ref{01}) -  (\ref{05}) can be linearized in the following form (Chandra and
Kumthekar, 2007):
\begin{eqnarray}
\frac{\partial \rho_1}{\partial t} + \rho_0 (\nabla . \stackrel{\rightarrow}
{v}_1 ) = 0 \hspace{3.5cm} \label{06}\\ 
\rho_0 \frac{\partial \hspace{-0.5mm} \stackrel{\rightarrow} v_1}{\partial t} =
-\nabla p_1 + \frac{1}{4 \pi} (\nabla \times \stackrel{\rightarrow}{B_1}) \times
 \stackrel{ \rightarrow}{B_0}- \nabla . \Pi_0  \label{07}\\ 
\frac{\partial \hspace{-1mm} \stackrel{\rightarrow}{B_1}}{\partial t} = \nabla
\times (\stackrel{\rightarrow}{v_1} \times \stackrel{\rightarrow}{B_0})
\hspace{3.3cm} \label{08}\\ 
\frac{\partial  p_1}{\partial t}+ \gamma p_0 (\nabla . \stackrel{\rightarrow}
{v}_1) + (\gamma -1) \kappa_\parallel k_z^2 T_1 = 0 
\hspace{4mm} \label{09}\\
\frac{p_1}{p_0} = \frac{\rho_1}{\rho_0}+ \frac{T_1}{T_0}
\hspace{4.5cm} \label{10}
\end{eqnarray}
                                                                                
\noindent
For the perturbations that are proportional to $\mbox{exp} [i (\stackrel{
\rightarrow}{k} \hspace{-1mm} . \hspace{-1mm} \stackrel{\rightarrow}{r} - 
\omega t)]$, equations (\ref{06}) - (\ref{10}) reduce to the following 
equations:
\begin{eqnarray}
\omega \rho_1 - \rho_0(k_x v_{1 x} + k_z v_{1 z}) = 0 \hspace{6.3cm} 
\label{11}\\
\omega\rho_0 v_{1 x}-k_x p_1-\frac{B_0}{4\pi}(k_x B_{1 z}-k_z B_{1 x})+
\frac{i\eta_0}{3}(k_x^2 v_{1 x}-2k_x k_z v_{1 z}) = 0 \label{12}\\
\omega\rho_0 v_{1 y} +\frac{B_0}{4\pi}(k_z B_{1 y}) = 0\hspace{6.9cm}
\label{13}\\
\omega\rho_0 v_{1 z}-k_z p_1+\frac{i\eta_0}{3}(4k_z^2 v_{1 z}-2k_x k_z v_{1 x})
= 0 \hspace{3.6cm}\label{14}\\ 
\omega B_{1 x} + k_z B_0 v_{1 x} = 0 \hspace{7.6cm}\label{15}\\
\omega B_{1 y} + k_z B_0 v_{1 y} = 0 \hspace{7.6cm}\label{16}\\
\omega B_{1 z} - k_x B_0 v_{1 x} = 0 \hspace{7.6cm}\label{17}\\
i\omega p_1 - i \rho_0 c_s^2 (k_x v_{1 x} + k_z v_{1 z}) - (\gamma-1)
\kappa_\parallel k_z^2 T_1 =0 \hspace{2.9cm} \label{18}\\ 
\frac{p_1}{p_0}-\frac{\rho_1}{\rho_0 }-\frac{T_1}{T_0} = 0 \hspace{7.9cm}
\label{19}
\end{eqnarray}

\noindent
These equations (\ref{11}) - (\ref{19}) are the same as the equations (11) - 
(19)
of PD. (In equation (11) of PD, $\rho$ must be $\rho_1$.)
Equations (\ref{13}) and (\ref{16}) for the variables $v_{1 y}$ and $B_{1 y}$ 
are decoupled from the rest and describe the Alfv\'{e}n waves. The rest of the 
equations for $p_1$, $\rho_1$, $T_1$, $B _{1 x}$, $B_{1 z}$, $v_{1 x}$ and  
$v_{1 z}$ describe damped magnetohydrostatic waves. Now, on substituting 
$B_{1 x}$ and $B_{1 z}$ from equations (\ref{15}) and (\ref{17}) in equations 
(\ref{12}) and (\ref{14}), respectively, we get
\begin{eqnarray}
\Big(\omega^2\rho_0 + \frac{i \omega\eta_0}{3}k_x^2-v_A^2\rho_0 k^2\Big) v_{1 x}
- \frac{2 i \omega\eta_0 k_x k_z}{3} \ v_{1 z} - k_x \omega p_1 = 0 \label{20}
\end{eqnarray}

\noindent
and
\vspace{-3mm}
\begin{eqnarray}
\frac{2 i \eta_0 k_x k_z}{3} \  v_{1 x} - \Big(\omega\rho_0+\frac{4 i 
\eta_0}{3} k_z^2 \Big) v_{1 z} + k_z p_1 = 0 \hspace{2.2cm} \label{21}
\end{eqnarray}

\noindent
 When we eliminate $\rho_1$ and $T_1$ from equations (\ref{11}), (\ref{18}) and 
(\ref{19}), we get
\begin{eqnarray}
(c_0 p_0 k_x -i \rho_0 c_s^2 k_x \omega) v_{1 x} + (c_0 p_0 k_z - i \rho_0 
c_s^2 k_z\omega) v_{1 z} - (c_0 \omega - i\omega^2) p_1 = 0   \label{22}
\end{eqnarray}

\noindent
where $c_0 = (\gamma -1) \kappa_\parallel k_z^2 T_0/p_0$; $c^2_s = \gamma
p_0/\rho_0$; and $v^2_A = B^2_0/ 4\pi \rho_0$. Thus, the equations (\ref{20}), 
(\ref{21}) and (\ref{22}) are obtained from the equations which are the same
as of PD.

\section{Dispersion relation}

For convenience, let us express equations (\ref{20}) - (\ref{22}) as
\begin{eqnarray}
a_{11} v_{1 x} + a_{12} v_{1 z} + a_{13} p_1 = 0 \hspace{1cm} \label{201} \\
a_{21} v_{1 x} + a_{22} v_{1 z} + a_{23} p_1 = 0 \hspace{1cm}\label{202} \\
a_{31} v_{1 x} + a_{32} v_{1 z} + a_{33} p_1 = 0 \hspace{1cm}\label{203}
\end{eqnarray}

\noindent
where the coefficients $a$'s are:
\begin{eqnarray}
a_{11} = \Big(\omega^2\rho_0 + \frac{i \omega\eta_0}{3}k_x^2-v_A^2\rho_0
k^2\Big); \hspace{10mm} a_{12} = - \frac{2 i \omega\eta_0}{3}k_x k_z;
\hspace{10mm} a_{13} = - k_x \omega \nonumber\\
a_{21} = \frac{2 i \eta_0}{3}k_x k_z; \hspace{25mm} a_{22} = - \omega\rho_0 - 
\frac{4 i \eta_0}{3} k_z^2; \hspace{25mm} a_{23} = k_z \nonumber\\
a_{31} = c_0 p_0 k_x -i \rho_0 c_s^2 k_x \omega; \hspace{10mm} a_{32} = c_0 p_0 
k_z - i \rho_0 c_s^2 k_z\omega; \hspace{10mm} a_{33} = i\omega^2 - c_0 \omega 
\nonumber
\end{eqnarray}

\noindent
For a non-trivial solution of the set of equations (\ref{201}), (\ref{202}) and
(\ref{203}), we must have
\begin{eqnarray}
\begin{array}{ccc}
\left|
\begin{array}{lll}
a_{11} & a_{12} & a_{13} \\
a_{21} & a_{22} & a_{23} \\
a_{31} & a_{32} & a_{33} \\
\end{array}
\right|
&  \hspace{-1mm} \begin{array}{c}
= 0 \\  \\ \\
\end{array}\\
\end{array} \nonumber
\end{eqnarray}

\noindent
Expansion of this determinant and substitution of the values of $a$'s gives the 
fifth degree polynomial:
\begin{eqnarray}
\omega^5 + i A  \omega^4 - B \omega^3 - i C \omega^2 + D \omega + i E = 0
\label{204}
\end{eqnarray}

\noindent
where
\begin{eqnarray}
A = c_0 + \frac{\eta_0}{3 \rho_0} (k_x^2 + 4 k_z^2) \hspace{8.4cm} \nonumber \\
B = \frac{c_0 \eta_0}{3 \rho_0} (k_x^2 + 4 k_z^2) + (c_s^2 + v_A^2) k^2
\hspace{6.7cm} \nonumber \\
C = \frac{3\eta_0}{\rho_0} c_s^2 k_x^2 k_z^2 + \frac{c_0 p_0 k^2}{\rho_0} +
v_A^2 c_0 k^2 + \frac{4 \eta_0 v_A^2 k_z^2 k^2}{ 3 \rho_0}
\hspace{4.3cm} \nonumber \\
D = \frac{3 c_0 p_0 \eta_0 k_x^2 k_z^2}{\rho_0^2} + \frac{4 \eta_0 c_0 v_A^2
k_z^2 k^2}{3 \rho_0} + v_A^2 c_s^2 k_z^2 k^2 \hspace{4.9cm} \nonumber \\
E = v_A^2 c_0 p_0 k_z^2 k^2/\rho_0 \hspace{9.2cm}  \nonumber \nonumber
\end{eqnarray}

\noindent
It obviously shows that the dispersion relation is a fifth degree polynomial.
Now, question arises how PD are getting the sixth degree polynomial. It can be
understood in the following manner.

\section{How PD is getting sixth degree polynomial}

The dispersion relation of PD can be derived in the following manner:
On eliminating $p_1$ from equations (\ref{201})  and (\ref{203}), we get
\begin{eqnarray}
(a_{11} a_{33} - a_{31} a_{13}) v_{1 x} + (a_{12} a_{33} - a_{32} a_{13})
v_{1 z} =  0 \label{301}
\end{eqnarray}

\noindent
On eliminating $p_1$ from equations (\ref{202})  and (\ref{203}), we get
\begin{eqnarray}
(a_{21} a_{33} - a_{31} a_{23}) v_{1 x} + (a_{22} a_{33} - a_{32} a_{23})
v_{1 z} =  0 \label{302}
\end{eqnarray}

\noindent
From equations (\ref{301}) and (\ref{302}), we have
\begin{eqnarray}
(a_{11} a_{33} - a_{31} a_{13}) (a_{22} a_{33} - a_{32} a_{23}) = 
(a_{21} a_{33} - a_{31} a_{23}) (a_{12} a_{33} - a_{32} a_{13})
\label{303}
\end{eqnarray}

\noindent
Substitution of the values of $a$'s in equation (\ref{303}), gives the 
dispersion relation 
\begin{eqnarray}
\omega^6  + i A' \omega^5 - B' \omega^4 - i C' \omega^3 + D'
\omega^2 +i E'  \omega - F' = 0 \label{304}
\end{eqnarray}

\noindent
where
\begin{eqnarray}
A' = 2 c_0 + c_1; \hspace{5cm} B' = ( c_s^2 + v_A^2 ) k^2 + c_0 (2 c_1 + c_0);
 \hspace{4cm}\nonumber\\ \nonumber\\
C' = c_2 + c_0(k^2 (c_s^2 + 2 v_A^2 + \frac{p_0}{\rho_0} )+ c_0 c_1)
\hspace{1.3cm}D' = c_s^2 c_6 + c_0 ( c_3 + c_0 c_4 );
 \hspace{4.9cm} \nonumber\\ \nonumber\\
 E' = c_0 \Big[c_0 c_5 + c_6 ( c_s^2 + \frac{p_0}{\rho_0} )\Big];
\hspace{2.8cm} F' = c_0^2 c_6 p_0/\rho_0 \hspace{6.6cm} \nonumber
\end{eqnarray}

\noindent
and
\begin{eqnarray}
 c_1 =  \eta_0 (k_x^2 + 4  k_z^2)/3 \rho_0; \hspace{3.7cm}
 c_2 =  \eta_0 k_z^2 (4 v_A^2 k^2 + 9 c_s^2  k_x^2 )/3 \rho_0
\hspace{2cm}  \nonumber \\ \nonumber \\
 c_3 =  \frac{\eta_0 k_z^2}{3 \rho_0} \Big(8 v_A^2 k^2 + 9 \Big( c_s^2
 + \frac{p_0}{\rho_0} \Big) k_x^2 \Big) \hspace{2.7cm} c_4 = \Big( v_A^2 +
\frac{p_0}{\rho_0}\Big) k^2 \hspace{30mm} \nonumber \\ \nonumber \\
c_5 = \frac{\eta_0 k_z^2}{3 \rho_0} \Big(4 v_A^2 k^2 + \frac{9 p_0 k_x^2 }
{\rho_0} \Big) \hspace{3.2cm}  c_6 = v_A^2 k^2 k_z^2 \hspace{47mm}  \nonumber
\end{eqnarray}

\noindent
This is the dispersion relation derived by PD. Now, here we can show that DP
has introduced an addition root in this dispersion relation.

\section{Discussion}

It can be easily found that
\begin{eqnarray}
\omega^6  + i A' \omega^5 - B' \omega^4 - i C' \omega^3 + D'
\omega^2 +i E'  \omega - F' = (\omega + i c_0) \times \nonumber \\
(\omega^5 + i A  \omega^4 - B \omega^3 - i C \omega^2 + D \omega + i E)
\hspace{-1.5cm} \nonumber
\end{eqnarray}

\noindent
showing that five roots of both equations (\ref{204}) and (\ref{304}) are 
common. The additional root $\omega = - i c_0$ is introduced by PD.  The five 
common roots, are of 
the form $i \alpha_1$, $- \beta_2 + i \alpha_2$, $\beta_2 + 
i \alpha_2$, $- \beta_3 + i \alpha_3$ and $\beta_3 + i \alpha_3$. The first root
corresponds to the thermal motion whereas the rest four give the slow-mode and
fast-mode waves. The sixth root $- i c_0$ also corresponds to the thermal 
motion. This has no significance, as we are interested only in the slow-mode 
and fast-mode waves. It is difficult to understand how science is affected when
one considers the sixth degree polynomial derived by PD.

It can finally be concluded that the dispersion relation is of fifth degree
polynomial. PD
got the sixth degree polynomial, as they have introduced an additional root.
Moreover, only five roots are sufficient for propagation of slow-mode and 
fast-mode waves.

\section*{Acknowledgments}

We are thankful to the DST, New Delhi and ISRO, Bangalore for financial support 
in the form of research projects. A part of this 
work was done during the visit to the IUCAA, Pune. Financial support from the 
IUCAA, Pune is thankfully acknowledged.

\end{document}